# Study on P-Type Doping of Mid-Wave and Long-Wave Infrared Mercury Cadmium Telluride

Arash Dehzangi[1,2], John Armstrong[1], Chris Schaake[1], Mark Skokan[1], Tyler Morrison[1], Justin Wilks[1], Sameer Ajmera[1], Mike Kinch[1]

[1]DRS Network & Imaging Systems, LLC', Dallas, TX, USA.
[2] Department of Electrical and Computer Engineering, University of Texas at Dallas, Texas, USA

We present in depth study of p-type doping concentration of mid-wave infrared (MWIR) and long-wave infrared (LWIR) mercury cadmium Telluride (HgCdTe) thin films. Annealing time was changed under specific conditions to achieve a stable copper (Cu) doping concentration for HgCdTe thin films. Both MWIR and LWIR HgCdTe material were grown by molecular beam epitaxy (MBE), where different trends were observed between LWIR and MWIR HgCdTe thin films by increasing anneal time. We also report the impact of different thickness (4μm, 6μm and 9μm) along with annealing time on doping level of LWIR HgCdTe thin films.

## INTRODUCTION:

Along with different material systems and technologies for LWIR and MWIR photodetection[1-9], mercury cadmium Telluride (HgCdTe) is exceptional material system with a broad range of military and commercial applications including photovoltaic devices, medical imaging, and infrared imaging and cameras. HgCdTe has a diverse detection capability due to its tunable bandgap spanning over the entire infrared spectrum, from the shortwave infrared (SWIR) to the very long wave infrared (VLWIR) regions.[10]

For longer wavelengths and hence smaller band gap energy, suppressing thermally generated dark current for infrared photodetectors becomes increasingly challenging. [11, 12] Therefore, low doping concentration in the absorber region of a detector is critical to control the dark current by minimizing Auger processes, which is one of the main sources of dark current in HgCdTe detectors.[13]

Controlling and tuning the doping concentration in the absorption region is a requirement to deliver high performance HgCdTe based focal plane arrays. This is important for both *n-on-p* and *p-on-n* HgCdTe photodiode architectures. For the case of *p-on-n* devices, controlling the *n-type* carrier concentration is achieved by using an extrinsic doping approach.[13] For *n-on-p* devices on the other hand, it would be much harder when it comes to control of *p-type* carrier concentration (as the absorption region) at low levels.[14, 15] For a *p-type* absorption region, a well-controlled area of electrically active *p-type* dopant with lower carrier concentration is essential to achieve longer minority-carrier lifetimes and hence higher performance devices.

The major acceptor for HgCdTe bulk material is found to be Li, Cu, Ag, P, Au and As. [16-18] While As is the most implemented acceptor dopant in industry and mostly introduced by ion implantation, [19-24] several studies have been reported on *p-type* doping using Cu, Ag and Au for both CdTe and HgCdTe via different techniques.[25-31] Work using a combination of interstitial-vacancy and mass spectroscopy models were proposed for Cu and Ag impurities. [32, 33] Cu is a fast diffusing dopant and was found to be completely active after post doping annealing compared to Ag and Au which were reported to be 50% and 10% active right after annealing. [34] The fundamental mechanisms behind *p-type* doping in HgCdTe or CdTe is still not fully clear [14, 35]. Further studies still would be needed to better comprehend issues such as low solubility of dopants due to the formation of secondary phases and self-compensation [15, 36]. For the case of Cu dopants, it has been demonstrated that after proper annealing it can segregate effectively from its original substrate/source thorough the HgCdTe epilayer absorption region.[37]

In this study we present the impact of annealing time on Cu-doped long-wave (LWIR) and mid-wave (MWIR) infrared HgCdTe thin films grown by MBE. The principal goal of the annealing is to achieve desirable *p-type* doping concentrations for Cu dopant in HgCdTe thin films. We present an empirical study with the focus on annealing time for both MWIR and LWIR HgCdTe thin films to compare the impact of annealing time on doping level. This study was designed to give an empirical mapping on Cu doped HgCdTe material grown by MBE under different anneal times.

## EXPERIMENTAL PROCEDURE:

In this work, the LWIR and MWIR $Hg_{1-x}Cd_xTe$ material were grown in a Tellurium-rich pressure by molecular beam epitaxy (MBE) on lattice matched $Cd_{0.96}Zn_{0.04}Te$ (211) substrates. The growth temperature was kept low (185-190°C) to provide epitaxial quality interfaces. The surface temperature was monitored by an infrared pyrometer. The Hg/Te flux ratio was optimized to provide monocrystalline HgCdTe growth and was monitored by using in-situ reflection high-energy electron diffraction (RHEED) analysis. The x values for LWIR and MWIR material were chosen to be ~0.23 and ~0.30 aiming for ~ 9.0-10.0 μm and 5.0-6.0 μm cut-off wavelength, respectively. More details about the growth were reported elsewhere. [38, 39]. The MWIR HgCdTe films were grown with 6 μm absorber thickness, while LWIR HgCdTe wafers were grown to 4, 6 and 9 μm thicknesses. The surface of all grown wafers was in-situ passivated by growing a 0.4 μm thick layer of wider bandgap HgCdTe at the top of the surface. This HgCdTe cap layer also acts as surface protection. The HgCdTe layer for both MWIR and LWIR material is low doped *n-type* during the growth with indium as the donor. All the wafers were annealed at 250 °C under saturated Hg pressure after growth, eliminating metal vacancies formed during growth and turning the LWIR material into *n-type* ($N_D > 1 \times 10^{14}$ cm$^{-3}$).

For *p-type* doping, copper-doped ZnS (ZnS:Cu) was sputter-deposited on the HgCdTe wafers as a diffusion source for copper atoms. The copper was diffused into the HgCdTe in a 250 °C anneal under $N_2$, equivalent to annealing under tellurium-rich conditions. After annealing, the distribution of the Cu dopants is assumed to be uniform throughout the HgCdTe samples. Hg out-diffusion during annealing was neglected due to presence of the ZnS deposited layer. [40] The anneal time was varied with times of 1, 2, 4 and 8 days. Hall measurements were performed on the copper-doped wafers at two different magnetic fields (0.2 T and 0.7 T).

## RESULTS AND DISCUSSION:

Fig 1a,b presents the acceptor doping concentration ($N_A$) for MWIR and LWIR HgCdTe with different annealing times with a 600 Å ZnS:Cu source. To create a baseline for the study, 600 Å Cu doped ZnS was chosen as the p-type doping source and varied in follow on experiments. The impact of Cu doped ZnS thickness on $N_A$ level will also be addressed later. As is shown in Fig 1, $N_A$ values increased with anneal times up to four days, however, the MWIR and LWIR samples

showed different behavior beyond four days. While the $N_A$ value increase continues for MWIR samples (Fig 1a), the $N_A$ value for LWIR wafers (Fig 1b) maximizes at 4 days and then almost saturates at longer annealing time.

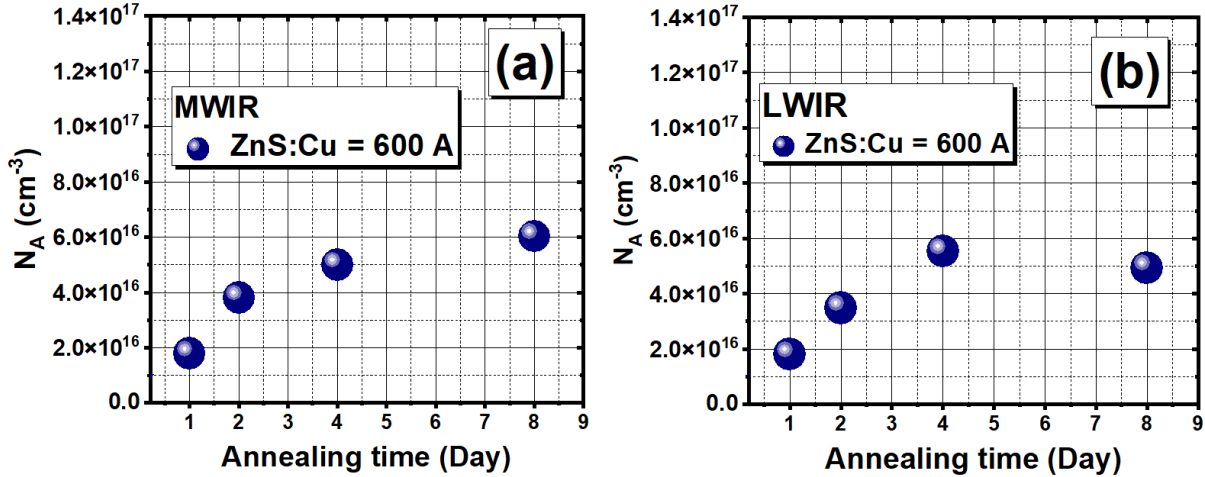

Fig 1. $N_A$ values vs annealing time measured at 77 K for (a) MWIR and (b) LWIR HgCdTe, the thickness of the HgCdTe films was 6 μm and annealing temperature was 250 °C

To give a better understanding of the impact of annealing, samples with no Cu dopant from MWIR and LWIR HgCdTe epilayers went through the same anneal process at 250 °C. An undoped ZnS layer was deposited on top of the epilayers to prevent surface decomposition in an identical condition with other samples. After 4 days anneal, the hole mobility values of the HgCdTe epilayers and associated *p-type* doping concentration are given in Table 1.

After a 250 °C tellurium-rich anneal, the HgCdTe crystals adjust their stoichiometry and metal vacancy concentration, leading to a change in doping from *n-type* (donor) to *p-type* (acceptor). Similar studies revealed that the vacancy concentration after annealing and the state of equilibrium kinetics is determined primarily by the annealing temperature.[41]

Table 1. mobility and $N_A$ values of the undoped HgCdTe wafers after four days anneal

| HgCdTe Material | Hole Mobility (cm²/Vs) 77 K | $N_A$ (1/cm³) |
|---|---|---|
| LWIR | 528 | $1.58 \times 10^{16}$ |
| MWIR | 412 | $1.37 \times 10^{16}$ |

the $N_A$ values presented in Table 1 show that the four days anneal for undoped HgCdTe generated approximately $1 \times 10^{16}$ cm$^{-3}$ Hg vacancies which behave as acceptors. The vacancy level for LWIR epilayers is a bit higher than MWIR but still in the same range. These vacancy levels values measured for these epilayers are in accordance with previous studies for vacancy equilibrium concentration at given temperatures. [16, 41, 42]

Given the values presented in Fig 1 for $N_A$ at different annealing times, a one day anneal seems to be ineffective on raising the doping level of the HgCdTe for both MWIR and LWIR, where the doping level remains in the range of metal vacancy doped material. We can conclude that $N_A$ values after one day anneal is in the level of Hg vacancies, and one day anneal is unable to introduce enough Cu dopants in the HgCdTe epilayers. The Cu doping increases in effect after two days of annealing and increases with additional anneal time.

Based on preliminary results, our study showed the ZnS:Cu thickness might have a strong influence on the $N_A$ values. To verify this assumption, additional doping tests were also performed on both MWIR and LWIR HgCdTe material using 800 Å and 1000 Å ZnS:Cu thickness.

Previous studies showed that Cu segregation is taking place in the HgCdTe layer and a gettering mechanism is associated with the presence of metal vacancies (mostly Hg). It has been suggested that the doping mechanism is dominated by interaction of interstitial copper with Hg vacancies and Cu segregation can be impacted by the scale of the Hg vacancy concentration. [37]

The summary of the $N_A$ result for different ZnS:Cu thickness is presented in Fig 2a,b for different annealing time. At first look, it confirms the trend that was observed in Fig 1. A linear trend exists for doping level with increasing ZnS:Cu thickness, since the thicker layer provides more source of Cu to diffuse into the bottom HgCdTe layer. For 800 and 1000 Å ZnS:Cu layers, the doping level rises close to $10^{+17}$ cm$^{-3}$ value range.

As expected, longer annealing time results in higher $N_A$ values for all ZnS thicknesses for both MWIR and LWIR material. For LWIR HgCdTe, a similar trend as presented in Fig 1 maintains and the doping concentration reaches saturation after 4 days annealing, whereas for MWIR material the trend continues up to 8 days, except that the rate of increase diminishes which is the sign of saturation. It can be inferred that the saturation state for $N_A$ values occurs at longer annealing time for MWIR material.

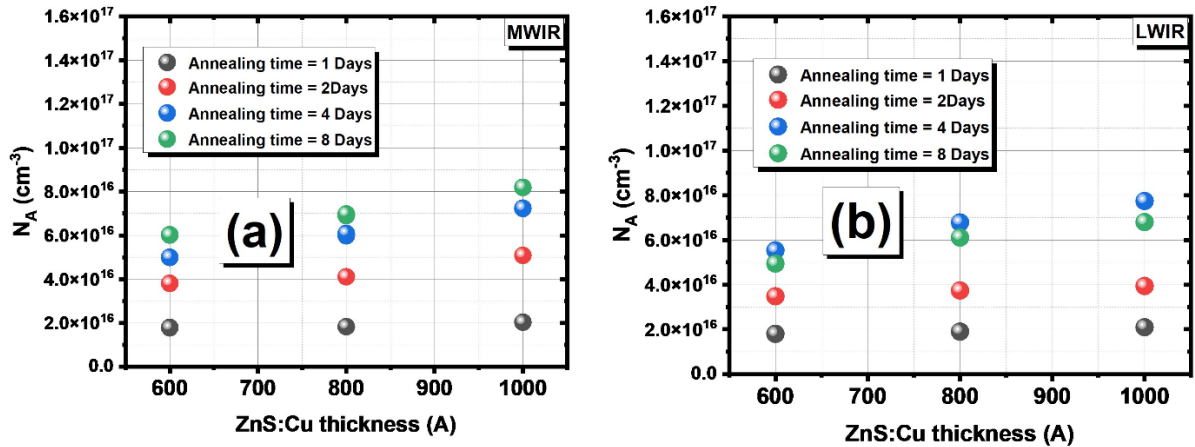

Fig 2. $N_A$ values vs ZnS:Cu thickness at 77 K for (a) MWIR and (b) LWIR HgCdTe, the thickness of the HgCdTe films were 6 μm and annealing temperature was 250 °C

The hole mobility comparison for the samples at different annealing time is also presented in Fig 3 for 6 μm thick LWIR and MWIR HgCdTe films. Two trends can be seen here: one is with increasing annealing time, and the second is with ZnS:Cu source thickness. For both MWIR and LWIR material, both trends result in raising the doping level in bulk HgCdTe and decreasing the mobility values, due to the impact of scattering.

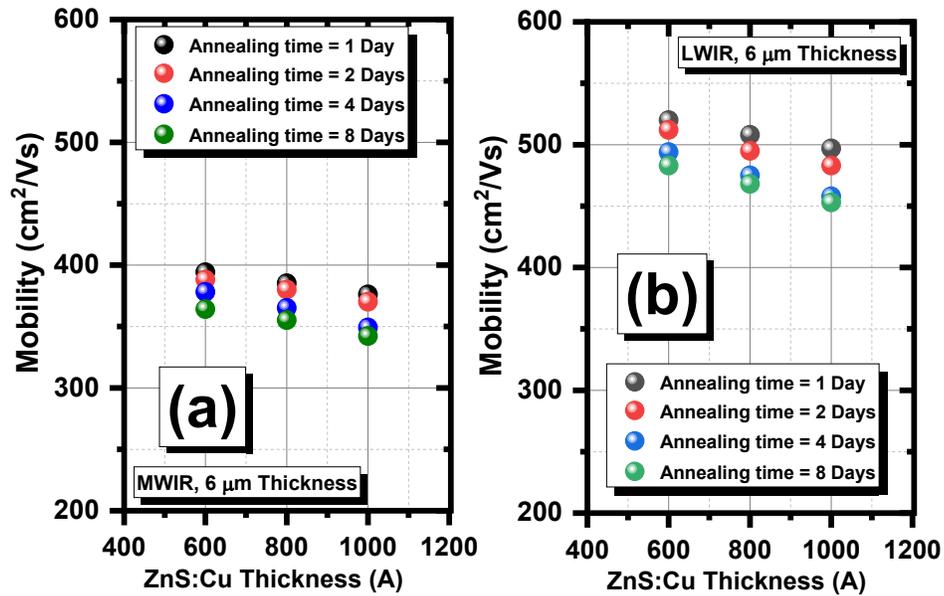

Fig 3. Mobility values vs ZnS:Cu thickness at 77 K for (a) MWIR and (b) LWIR HgCdTe, the thickness of the HgCdTe films were 6 μm and annealing temperature was 250 °C

As expected, the mobility values are lower for MWIR material compared to LWIR HgCdTe, with the lowest value of 342 cm$^2$/Vs associated with the highest $N_A$ value of $8.19\times10^{16}$ cm$^{-3}$ for 8 days anneal with ZnS:Cu thickness of 1000 Å. For LWIR material the mobility values for 4 and 8 days annealing fall in the same level due to saturation of doping level observed for those annealing times (Fig 1 and Fig 2). For LWIR HgCdTe the lowest value of mobility was 453 cm$^2$/Vs corresponding to $N_A$ value of $6.8\times10^{16}$ cm$^{-3}$ for 4 days anneal.

To give better perspective on the relationship of mobility, annealing and doping concentration, Fig 4a presents the carrier mobilities for both LWIR and MWIR HgCdTe as a function of doping concentration from the measured data. Different annealing times were also highlighted to follow the trend with annealing time. For each anneal time, three mobility values are shown which are associated with each ZnS:Cu thickness. To give some sense of comparison, in Fig 4a, the mobility values and corresponding doping concentration for undoped (4 days anneal) LWIR and MWIR HgCdTe are shown with a 'star' symbol (also shown in Table 1). The trend of decreasing mobility with increasing doping is expected due to the scattering of the carriers by doping atoms.[43]

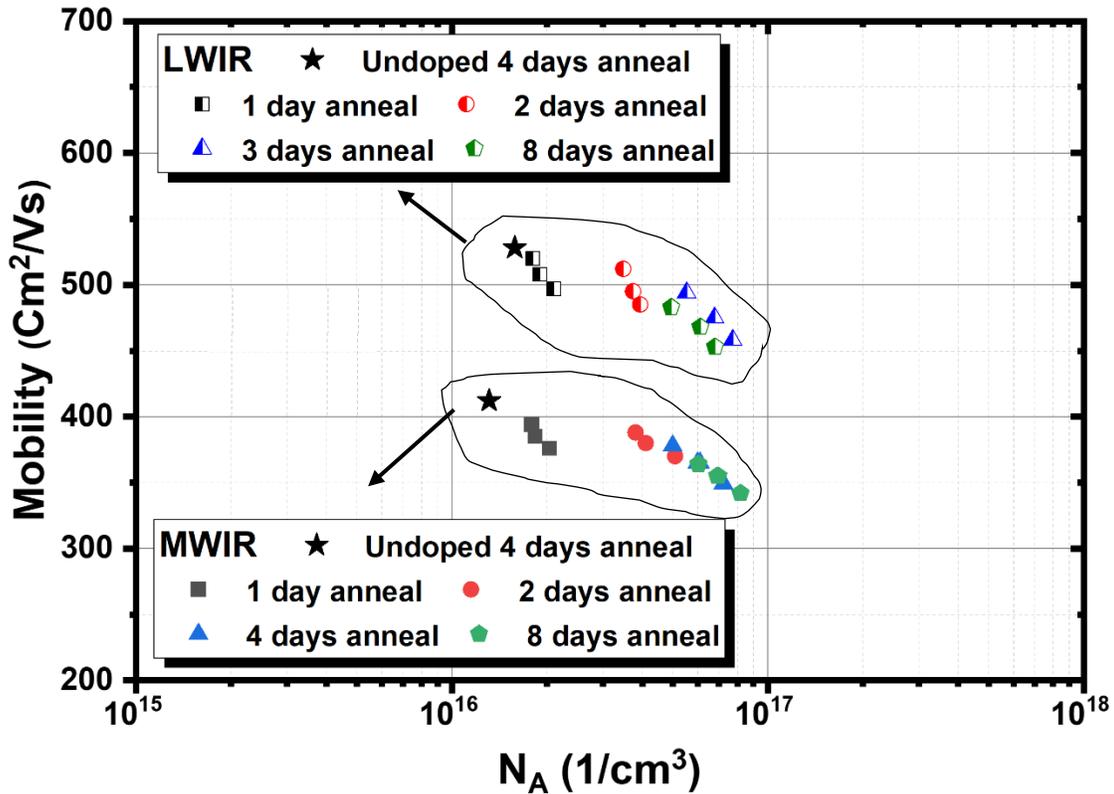

Fig 4. a) Mobility values vs $N_A$ at 77 K for LWIR and MWIR HgCdTe at different annealing time, the thickness of the HgCdTe films was 6 μm and annealing temperature was 250 °C. The mobility values for undoped LWIR and MWIR HgCdTe after 4 days anneal shown with star symbol.

For LWIR infrared photodetectors, in order to achieve a high absorption quantum efficiency, the absorber should be thicker than those typically used for MWIR detectors. [13, 44] Thus, various thicknesses for LWIR HgCdTe thin films were explored for varying anneal time and ZnS:Cu thickness. HgCdTe LWIR wafers with thickness of 4 and 9 μm were grown by MBE in addition to the 6 μm thick HgCdTe wafers previously mentioned. We also performed the same study on different LWIR material with different thickness. In all cases, we found the same trend observed for 6 μm thickness LWIR HgCdTe and doping level was saturated after 4 days anneal. The result for 4 days annealing time is shown in Fig 5. Fig 5 demonstrates the $N_A$ values versus 1/ HgCdTe thickness at 77 K for LWIR material with different ZnS:Cu thicknesses of 600, 800 and 1000A. The result reveals an inverse trend between the thickness and doping level for the LWIR HgCdTe. By increasing the thickness of HgCdTe layer and having larger bulk volume for certain ZnS:Cu layer thicknesses (as the source of Cu dopant), a drop in the $N_A$ values occurs. This is expected, since the source of dopants was kept constant, and the volume of the bulk material increases. The other trend in Fig 5 is that by increasing the thickness of ZnS:Cu layer and providing higher source of Cu dopants, the p-type doping level increases for all the LWIR HgCdTe with different thicknesses. The trends are crucial toward optimizing the doping concentration for different thickness of LWIR HgCdTe for different applications.

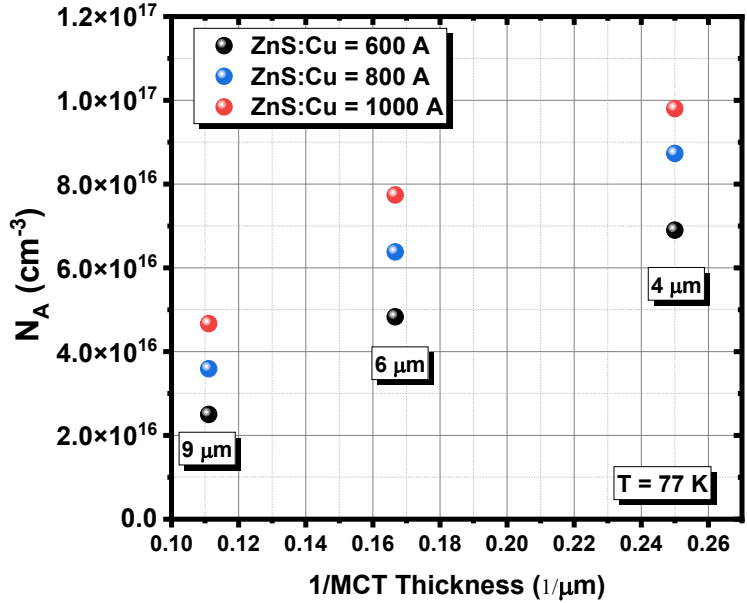

Fig 5. a) $N_A$ values vs 1/ HgCdTe thickness at 77 K for LWIR material at 4 days annealing time (at 250 °C) with different ZnS:Cu thickness, the HgCdTe film thickness were 4, 6 and 9 μm.

The carrier mobilities for different thicknesses of LWIR HgCdTe as a function of doping concentration are shown in Fig 6. The data represents the result after 4 days anneal for LWIR HgCdTe, with three mobility values for each HgCdTe thickness, which are associated to three ZnS:Cu thickness (600, 800 and 1000 Å). Similar trend with Fig 4 can be observed where a drop in mobility values exist by increasing the doping level.

It is worth noting that mobility is only impacted by doping concentration and not the thickness of the HgCdTe layer. After 4 days annealing and 1000 Å ZnS:Cu thickness, the mobility value for HgCdTe with 4 μm reaches its lowest value 414 cm$^2$/Vs corresponding to $N_A$ value of $9.8 \times 10^{16}$ cm$^{-3}$. For the same condition, the mobility value for HgCdTe with 9 μm reaches its lowest value 501 cm$^2$/Vs corresponding to $N_A$ value of $4.67 \times 10^{16}$ cm$^{-3}$.

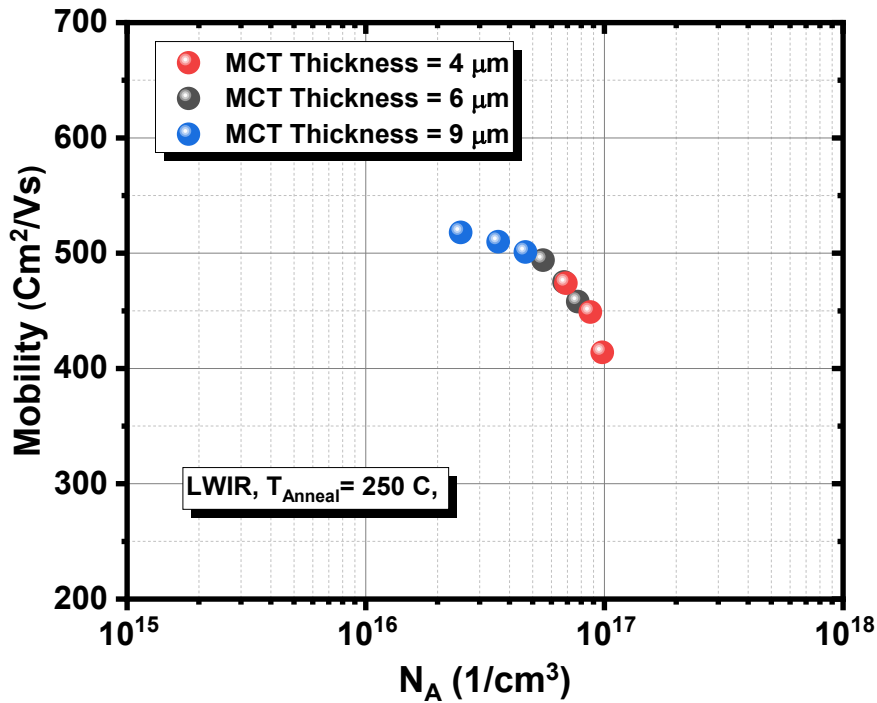

Fig 6. Mobility values vs $N_A$ at 77 K for LWIR HgCdTe with different HgCdTe film thickness, the thickness of the HgCdTe was 4, 6 and 9 μm. Annealing time was 4 days for all the samples annealed at 250 °C.

## CONCLUSIONS:

In conclusion, the impact of annealing time on Cu-doped MWIR and LWIR infrared HgCdTe thin films grown by MBE was studied and compared. Longer annealing time leads to higher doping concentration and lower mobility values for both LWIR and MWIR material. However, a different behavior was observed between LWIR and MWIR HgCdTe thin films when it comes to longer annealing times. The saturation time for MWIR HgCdTe is longer compared to LWIR HgCdTe. The reason for this trend is unknown and needs further investigation. Different values of Cu dopants were also studied to examine affects on doping concentration. The same experiment was run on LWIR HgCdTe films with different thickness where a linear trend between the p-type doping concentration and HgCdTe thickness was observed. This study was designed to give an empirical mapping on Cu doped HgCdTe material grown by MBE under different anneal times.